\documentstyle[epsfig,12pt]{article}

\newcommand{\menor}{\mbox{\raisebox{-0.4ex}
{$\;\stackrel{<}{\scriptstyle \sim}\;$}}}
\def\etjet{E_T^{jet}}
\def\etajet{\eta^{jet}}

\def\etaphi{\eta-\varphi}

\def\g2{GeV$^2$}
\def\q2{Q^2}
\def\oalphas3{O(\alpha_S^3)}

\def\etjb{E^{jet}_{T}({\rm Breit})}
\def\etajl{\eta^{jet}({\rm Lab})}
\def\etajb{\eta^{jet}({\rm Breit})}
\begin{document}

\rightline{FTUAM-99/07}
\rightline{March 1999}
 
\vspace{0.5 cm} 
\begin{center}
{\Large {\bf Jet shapes in $ep$ collisions at HERA\footnote{Plenary
talk given at the 3rd UK Phenomenology Workshop on HERA Physics,
Durham, UK, September 1998.}\\}}
\end{center}
    
\begin{center}
   {J. Terron (on behalf of the H1 and ZEUS collaborations)}
\end{center}
\begin{center}
   {\it Universidad Aut\' onoma de Madrid, 28049 Madrid, Spain}
\end{center}
 
\begin{abstract}
New measurements of the jet shape in $ep$ collisions at HERA using the
$k_T$-cluster jet algorithm are presented. 
\end{abstract}

\section{Introduction}
 The investigation of the internal structure of jets gives insight into 
the transition between a parton produced in a hard process and the
experimentally observable spray of hadrons. The internal structure of a
jet is expected to depend mainly on the type of primary parton, quark or
gluon, from which it originated and to a lesser extent on the particular
hard scattering process. A useful representation of the jet's internal
structure is given by the jet shape \cite{sdellis}. At sufficiently high
jet energy, where fragmentation effects become negligible, the jet shape
should be calculable by perturbative QCD. Measurements of the jet shape
provide a stringent test of pQCD calculations beyond leading order.
Gluon jets are predicted to be broader than quark jets due to the larger
colour charge of the gluon. The dependence of the structure of quark and
gluon jets on the production process can be investigated by comparing
measurements of the jet shape in different reactions in which the 
final-state jets are predominantly quark or gluon initiated.

 Measurements of the jet shape were made in $\bar{p}p$ collisions at
$\sqrt{s}=1.8$~TeV \cite{cdf1,d01} and in $e^+e^-$ interactions at LEP1 
\cite{opal1}. It was observed \cite{opal1} that the jets in $e^+e^-$ are
significantly narrower than those in $\bar{p}p$ and most of this 
difference was ascribed to the different mixtures of quark and gluon 
jets in the two production processes. At HERA, measurements have been 
presented of the jet shape in quasi-real photon proton collisions 
(photoproduction) \cite{zshape97} and in neutral- and charged-current 
deep inelastic scattering (DIS) \cite{zshape98}. In photoproduction, the 
jets were observed to become broader as the jet pseudorapidity 
($\eta^{jet}$) increases in agreement with the predicted increase in the
fraction of final-state gluon jets. In DIS, the jet shapes in neutral- 
and charged-current processes were found to be very similar. The jet 
shapes in DIS were observed to be similar to those in $e^+e^-$ 
interactions and narrower than those in $\bar{p}p$ collisions. Since the
jets in $e^+e^-$ interactions and $e^+p$ DIS are predominantly quark 
initiated in both cases, the similarity in the jet shapes indicates that 
the pattern of QCD radiation within a quark jet is to a large
extent independent of the hard scattering process in these reactions.

 New measurements of the jet shape using the $k_T$-cluster algorithm
\cite{ktcl1,ktcl2} in photoproduction \cite{juan} and DIS 
\cite{hshape98} at HERA provide an improved test of pQCD calculations 
\cite{seymour} and are presented here. During 1994-1997 HERA operated 
with positrons of energy $E_e=27.5$~GeV colliding with protons of energy
$E_p=820$~GeV.

\section{Measurement of the jet shape in photoproduction}
 At HERA, quasi-real photon proton collisions are studied via $ep$ 
scattering at low four-momentum transfers ($Q^2 \approx 0$, where $Q^2$ 
is the virtuality of the exchanged photon). Jets are searched for in the
pseudorapidity ($\eta$) - azimuth ($\varphi$) plane of the laboratory 
frame using the inclusive $k_T$-cluster algorithm \cite{ktcl2} (see 
\cite{ktcl3} for the experimental implementation). The jet variables are
defined according to the Snowmass convention \cite{snowmass}. The 
inclusive sample of jets with transverse energy $\etjet > 17$~GeV and 
$-1 < \etajet < 2$ has been studied.
 
 The differential jet shape is defined as the average fraction of the
jet's transverse energy that lies inside an annulus in the $\etaphi$ 
plane of inner (outer) radius $r-\Delta r/2$ ($r+\Delta r/2$) concentric
with the jet axis \cite{sdellis}:
\begin{equation}
 \rho(r) \equiv \frac{1}{N_{jets}} \frac{1}{\Delta r}
 \sum_{jets}\frac{E_T(r-\Delta r/2,r+\Delta r/2)}{E_T(0,1)},
\end{equation}
where $E_T(r-\Delta r/2,r+\Delta r/2)$ is the transverse energy within 
the given annulus and $N_{jets}$ is the total number of jets in the 
sample. The differential jet shape has been measured for $r$ values 
varying from $0.05$ to $0.95$ in $\Delta r=0.1$ increments.

\begin{figure}
\vspace{1.5cm}
\centerline{\mbox{
\epsfig{figure=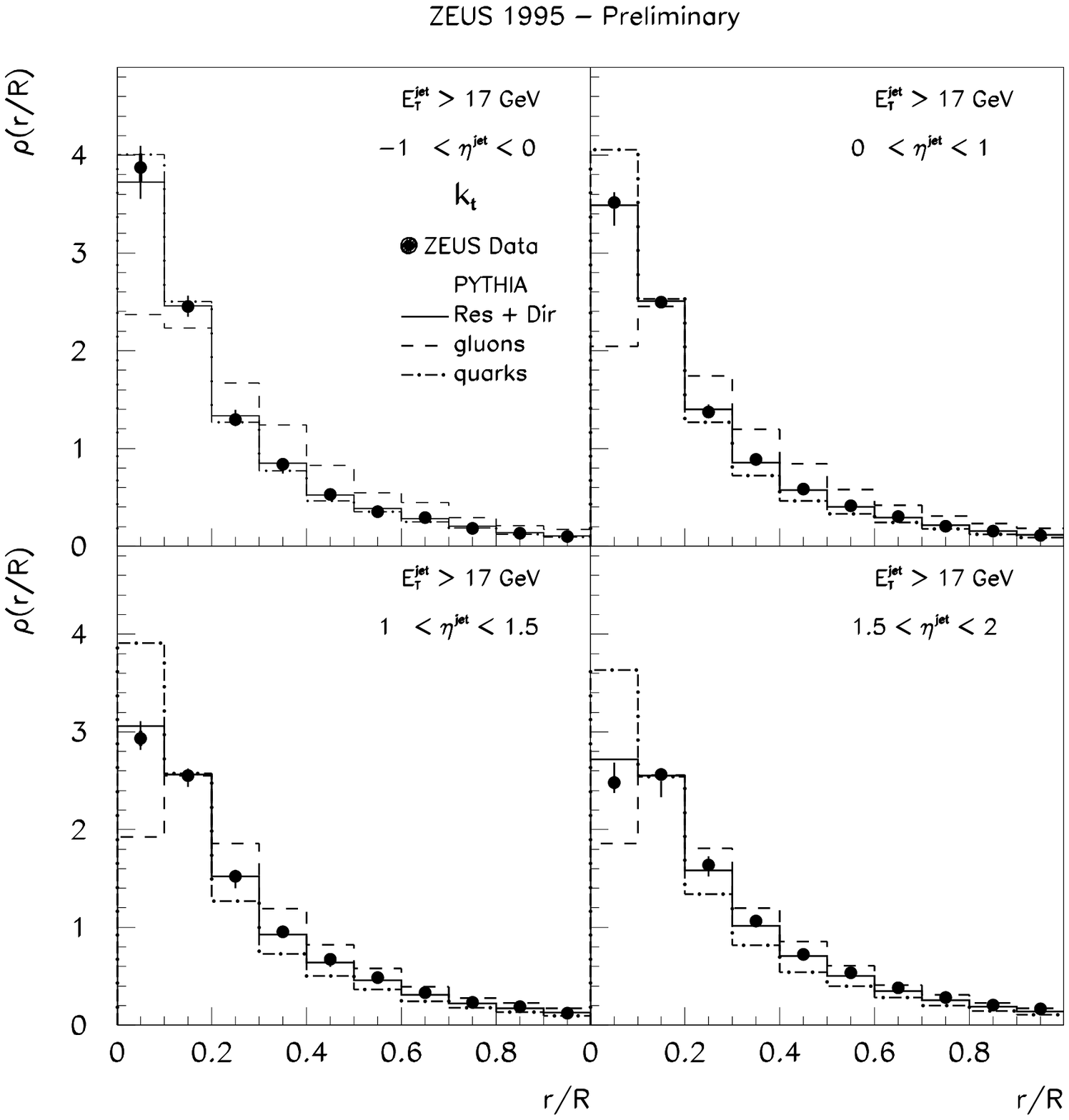,width=8cm}
\hspace{-1cm}
\epsfig{figure=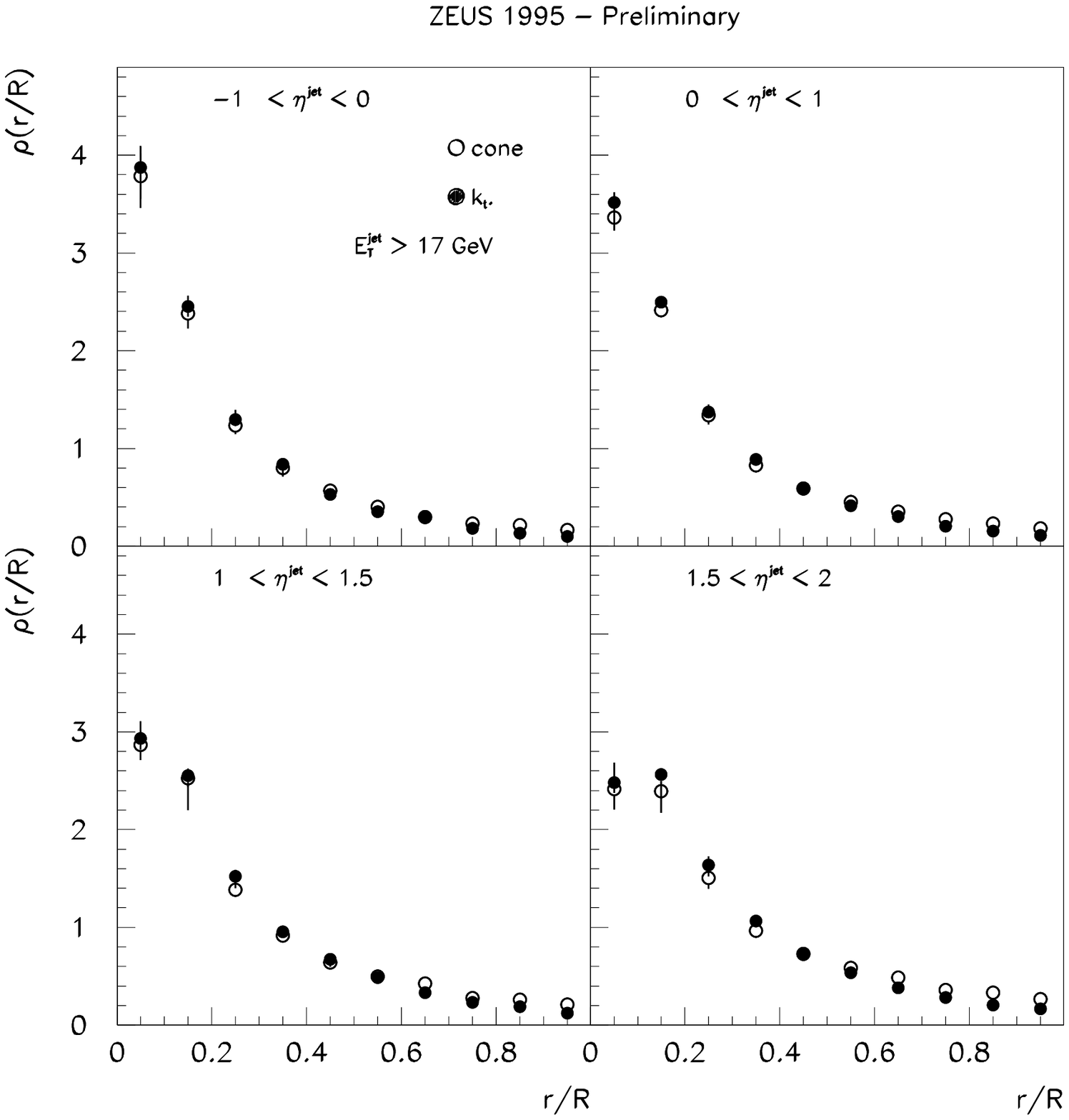,width=8cm}}}
\vspace{-2.5cm}
\caption{Measured differential jet shapes corrected to the hadron level,
 $\rho(r)$, using the $k_T$-cluster algorithm for jets with 
$\etjet > 17$~GeV in different $\etajet$ regions (black dots). For 
comparison, the predictions of PYTHIA for photoproduced jets (solid 
histogram) and separately for quark and gluon jets are shown in the 
left-hand side of the figure. The measurements using the cone algorithm 
with $R=1$ (open circles) are compared to those with
the $k_T$-cluster algorithm in the right-hand side of the figure.
\label{zeuskt}}
\end{figure}

 The differential jet shape has been measured using the ZEUS
uranium-scintillator calorimeter and corrected to the hadron level. The
measurements are given in the kinematic region defined by 
$Q^2 < 1$~GeV$^2$ (with a median of $Q^2 \approx 10^{-3}$) and 
photon-proton centre-of-mass energies between 134 and 277~GeV. The 
measured differential jet shapes \cite{juan} for different regions in 
$\etajet$ are shown in Figure~\ref{zeuskt} (black dots). It is observed 
that the jet broadens as $\etajet$ increases in agreement with our 
previous observation using an iterative cone algorithm with radius 
$R=1$ \cite{zshape97}.

 The predicted jet shapes at the hadron level from a leading-logarithm
parton-shower Monte Carlo calculation using PYTHIA \cite{pythia} are
compared to the measurements in the left-hand side of 
Figure~\ref{zeuskt}. The calculations include initial- and final-state 
parton radiation, and the fragmentation into hadrons is performed using 
the LUND string model. The measured jet shapes are found to be well 
described by the predictions (solid histogram). The jet shapes, as 
predicted by PYTHIA, for quark (dot-dashed histogram) and gluon (dashed 
histogram) jets are also shown in Figure~\ref{zeuskt}: the broadening of
the jets in the data as $\etajet$ increases is consistent with an 
increasing fraction of gluon jets.

 It has been shown \cite{seymour} that the inclusive $k_T$-cluster 
algorithm provides, at present, the best jet algorithm from the 
theoretical point of view since the problem of overlapping jets, which 
affects e.g. the iterative cone algorithm \cite{cone}, is avoided. To 
quantify the effects of the specific jet algorithm on the jet shape, the 
measurements have been repeated using the iterative cone algorithm with 
radius $R=1$. The results (open circles) are compared to those using the
$k_T$-cluster algorithm (black dots) in the right-hand side of 
Figure~\ref{zeuskt}: the measured jet shapes differ by less than 10\% in
the region $r<0.6$. For larger values of $r$ differences are expected 
since in the case of the iterative cone algorithm only those particles 
within a cone concentric to the jet axis are assigned to the jet while 
in the $k_T$ no such a restriction is imposed. Thus, in spite of the 
differences between the two algorithms the jet shapes are observed to be
very similar in the region $r<0.6$ and demand pQCD calculations which 
are able to reproduce the features of the specific jet algorithm with an
accuracy better than 10\%. Next-to-leading order QCD calculations of the
jet shape with the $k_T$-cluster algorithm, which are not available at 
present, are needed to meet such a requirement.

\section{Measurement of the jet shape in deep inelastic scattering}
 Measurements have been made of the internal jet structure in a sample of
inclusive dijet neutral-current DIS events \cite{hshape98},
$ e^+ \; + \; p \rightarrow e^+ \; + \; {\rm jet} \; + \; {\rm jet}
 \; + \; {\rm X}$,
in the kinematic region defined by $10<Q^2 \menor 120$~GeV$^2$ and
$2\cdot 10^{-4} \menor x_{Bj} \menor 8 \cdot 10^{-3}$. Jets are searched 
for in the $\etaphi$ plane of the Breit frame using the inclusive 
$k_T$-cluster algorithm \cite{ktcl2}. The jet variables are defined 
according to the Snowmass convention \cite{snowmass}. The sample
of inclusive dijet events with transverse energy (with respect to the
direction of the virtual photon in the Breit frame) $\etjb>5$~GeV and 
$-1 < \etajl < 2$ has been investigated.

 In this analysis the internal structure of a jet is studied in terms of
the integrated jet shape, $\Psi(r)$, which is defined as the average 
fraction of the jet's transverse energy that lies inside a subcone in 
the $\etaphi$ plane of radius $r$ concentric with the jet 
axis~\cite{sdellis}:
\begin{equation}
 \Psi(r) \equiv \frac{1}{N_{jets}}\sum_{jets} \frac{E_T(r)}{\etjb},
\end{equation}
where $E_T(r)$ is the transverse energy within the subcone of radius $r$
and $N_{jets}$ is the total number of jets in the sample.

\begin{figure}
\centerline{\psfig{figure=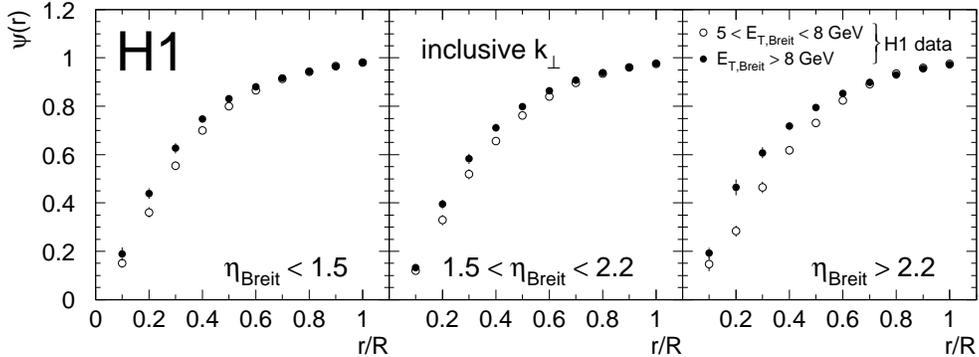,height=5cm}} 
\caption{Measurements of the integrated jet shape in the inclusive 
sample of dijet NC DIS events with $5<\etjb <8$~GeV (open circles) and 
$\etjb > 8$~GeV (black dots) for three regions in $\etajb$.
\label{h1diskt}}
\end{figure}

 The measured integrated jet shapes \cite{hshape98} are shown in
Figure~\ref{h1diskt} for two ranges in $\etjb$ and three regions in 
$\etajb$ (negative $\etajb$ corresponds to the virtual-photon 
hemisphere). The jets are observed to be more collimated as $\etjb$ 
increases. On the other hand, the jets become broader as $\etajb$ 
increases and this effect is more pronounced at lower $\etjb$. The 
measured dependence of the jet shape on $\etjb$ and $\etajb$ is roughly 
reproduced by the predictions of various QCD-based models (not shown 
here; see \cite{hshape98}). However, studies based on these models show 
that in the region of $\etjb$ considered in this analysis the jet shape 
is strongly influenced by hadronization. Thus, measurements with higher 
$\etjet$ (see \cite{zshape97,zshape98}) are needed to test pQCD 
calculations.

 The measurements have been repeated \cite{hshape98} using a version of 
the iterative cone algorithm \cite{cone3} which allows improved pQCD 
calculations of the jet shape. The measured jet shapes with the 
$k_T$-cluster and the iterative cone algorithms are observed to be very 
similar in the region $\etjb >8$~GeV and $\etajb<2.2$. For lower $\etjb$
or higher $\etajb$ the jets identified with the cone algorithm are 
broader. From this comparison and that in photoproduction (with 
$\etjet >17$~GeV), it is concluded that the effects of the specific jet 
algorithm decrease rapidly as $\etjet$ increases. The measurements of 
jet shapes with the $k_T$-cluster algorithm at high 
$\etjet$ ($\etjet > 17$~GeV) constitute a challenge to pQCD calculations.

\vspace{1cm}

{\bf Acknowledgements:} I would like to thank the organizers for the 
superb location of the conference. The help of my colleagues from 
the H1 and ZEUS Collaborations and, in particular, from Claudia 
Glasman, in the preparation of the material 
reported here is gratefully appreciated.


\end{document}